\documentclass[twocolumn,showpacs,amsmath,amssymb,aps,prd,nofootinbib]{revtex4}

\usepackage{graphicx}
\usepackage{color} 
\usepackage{soul}  

\begin{document}

\title{Gravitational wave generation in loop quantum cosmology}

\author{Paulo M. S\'a}

\email{pmsa@ualg.pt}

\affiliation{Departamento de F\'{\i}sica, Faculdade de Ci\^encias
e Tecnologia, Universidade do Algarve, Campus de Gambelas,
8005-139 Faro, Portugal}

\author{Alfredo B. Henriques}

\email{alfredo.henriques@ist.utl.pt}

\affiliation{Centro Multidisciplinar de Astrof\'{\i}sica -- CENTRA and
Departamento de F\'{\i}sica, Instituto Superior T\'ecnico, UTL, Av.\ Rovisco
Pais, 1049-001 Lisboa, Portugal}

\date{July 30, 2011}

\begin{abstract}
We calculate the full spectrum, as observed today, of the cosmological
gravitational waves generated within a model based on loop quantum cosmology.
It is assumed that the universe, after the transition to the classical regime,
undergoes a period of inflation driven by a scalar field with a chaotic-type
potential. Our analysis shows that, for certain conditions, loop quantum
effects leave a clear signature on the spectrum, namely, an over-production of
low-frequency gravitational waves. One of the aims of our work is to show that
loop quantum cosmology models can be tested and that, more generally,
pre-inflationary physical processes, contrary to what is usually assumed,
leave their imprint in those spectra and can also be tested.
\end{abstract}


\maketitle

\section{Introduction}

Although gravitational waves of cosmological origin have not yet been
detected, they are at present the object of a considerable research effort, as
they may provide us with a unique telescope to the earliest stages of the
formation of the universe. At the same time we also witness an increasing
interest in the applications of the ideas of loop quantum gravity (for a
review, see Ref.~\cite{rovelli-2008}) to the problems of cosmology, a field
known as loop quantum cosmology, interest that started after a series of
seminal papers by Bojowald \cite{bojowald-varios}, where a number of important
results were obtained, among them the possibility of removing in a natural way
the presence of the cosmological singularity (for a review about loop quantum
cosmology, see Ref.~\cite{bojowald-2008}).

Tests of loop quantum cosmology have already been proposed
\cite{bojowald-lidsey-mulryne-singh-tavakol,tsujikawa-singh-maartens,hossain,
copeland-mulryne-nunes-shaeri,bojowald-calcagni,bojowald-calcagni-tsujikawa-A,
bojowald-calcagni-tsujikawa-B}, showing that loop effects may appear, albeit
in an indirect way, on the cosmic microwave background radiation on the
largest scales. Loop quantum cosmology gives rise to changes in the dynamical
equations driving the expansion of the universe; these are connected with
modifications in the equation of state of the matter content of the universe,
which in turn result in the production of gravitational waves. These
gravitational waves are the focus of our work, where we show that they may
leave an important imprint in today's power spectrum. This at first may seem
surprising, as it has usually been assumed that inflation, by its
characteristics, among them the enormous increase in the scale of the
universe, would remove any kind of information coming from physical phenomena
taking place in pre-inflationary times. This is not the case. Pre-inflationary
physical features affect in a different way different frequencies and the
memory of such differences survives through the inflationary period, and can
be read today in the power spectrum.

Our paper is organized as follows. In the next section we describe the loop
quantum cosmology model used in our work and write the equations of motion for
the semiclassical and classical stages of evolution of the universe. Taking
into account the constraints arising from measurements of the cosmic microwave
background radiation, we specify the values of the parameters and the initial
conditions and solve numerically the evolutionary equations, from the
semiclassical pre-inflationary epoch till the present time. In
Section~\ref{sect-gravitational-waves} we calculate the full
gravitational-wave spectrum, for frequencies ranging from about
$10^{-17}$~rad/s to about $10^9$~rad/s, using the method of continuous
Bogoliubov coefficients. The influence on the spectrum of the ambiguity
parameters of loop quantum cosmology is carefully analyzed. In
section~\ref{sect-conclusions} we summarize the results we have obtained.

\vspace{-2mm}
\section{The evolution of the universe\label{section-3}}

The evolution of the universe is naturally divided in two stages,
corresponding to the semiclassical and the classical regimes. During the
semiclassical stage of evolution, the equations of standard cosmology have to
be modified in order to account for loop quantum effects. After a few Planck
times of evolution, the transition between the semiclassical and classical
regimes takes place and the universe further evolves according to the standard
cosmological model, namely, undergoes a period of inflation, driven by a
scalar field $\phi$ with chaotic-type potential
\begin{eqnarray}
 V(\phi)=\frac12 m_\phi^2\phi^2,
\end{eqnarray}
followed by reheating and, successively, by radiation-dominated,
matter-dominated and dark energy-dominated periods.

During the semiclassical stage, the modified evolutionary equations for the
scale factor $a$ and the scalar field $\phi$ are given by
\cite{tsujikawa-singh-maartens,mielczarek-szydlowski,grain-et-al}
\begin{eqnarray}
\left( \frac{\dot{a}}{a} \right)^2 = \frac{8\pi}{3m_{\textsc{P}}^2} \left[
\frac{\dot{\phi}^2}{2D(q)} + V(\phi) \right], \label{H}
\end{eqnarray}
\vspace{-5mm}
\begin{eqnarray}
\frac{\ddot{a}}{a} = \frac{8\pi}{3m_{\textsc{P}}^2} \left[  V(\phi)-
\frac{\dot{\phi}^2}{D(q)} \right] + \frac{2\pi \dot{\phi}^2}{m_{\textsc{P}}^2}
\frac{F(q)}{D(q)}, \label{raichaid}
\end{eqnarray}
\vspace{-5mm}
\begin{eqnarray}
\ddot{\phi} = -3\frac{\dot{a}}{a} \Big[ 1-F(q) \Big] \dot{\phi} -D (q)
\frac{\partial V}{\partial \phi} \label{phi},
\end{eqnarray}
where a flat Friedmann-Robertson-Walker background is assumed,
$m_{\textsc{P}}$ is the Planck mass, a dot denotes a derivative with respect
to the cosmic time $t$ and the following notation was introduced
\cite{bojowald-2004}
\begin{eqnarray}
D(q) &= & \left( \frac{3}{2\ell} \right)^{\frac{3}{2-2\ell}} q^{3/2} \left\{
\frac{1}{\ell+2} \Big[ (q+1)^{\ell+2}-|q-1|^{\ell+2} \Big] \nonumber \right.
\\
&& \hspace{-10mm} \left. -\frac{q}{\ell+1} \Big[ (q+1)^{\ell+1}- \mbox{sign}
(q-1) |q-1|^{\ell+1} \Big] \right\}^\frac{3}{2-2\ell},
\end{eqnarray}
and
\begin{eqnarray}
F(q) &=& \frac{1}{\ell-1} \Big\{ (\ell^2-1) \Big[
(q+1)^{\ell+2}-|q-1|^{\ell+2}\Big]  \nonumber
\\
&& \hspace{-10mm} -(2\ell-1)(\ell+2)q \Big[
(q+1)^{\ell+1}-\mbox{sign}(q-1)|q-1|^{\ell+1} \Big]
\nonumber \\
& &  \hspace{-10mm}
+(\ell+1)(\ell+2)q^2 \Big[ (q+1)^{\ell}-|q-1|^{\ell} \Big] \Big\}  \nonumber \\
&& \hspace{-10mm} \times \Big\{ (\ell+1) \Big[ (q+1)^{\ell+2}-|q-1|^{\ell+2} \Big] \nonumber  \\
&& \hspace{-10mm}  - (\ell+2) q \Big[ (q+1)^{\ell+1}- \mbox{sign} (q-1)
|q-1|^{\ell+1} \Big] \Big\}^{-1},
\end{eqnarray}
with\footnote{Different values for the Barbero-Immirzi parameter $\gamma$ can
be found in literature. We use the value obtained by Meissner from black-hole
entropy calculations \cite{meissner}.}
\begin{eqnarray}
q=\left( \frac{a}{a_*} \right)^2, \quad a_*=\sqrt{\frac{\gamma j}{3}}\,
\ell_{\textsc{P}}, \quad  \gamma=0.2375.
\end{eqnarray}
In the above expressions $\ell$ and $j$ are the so-called ambiguity parameters
and $\ell_{\textsc{P}}$ denotes the Planck length.

Equations (\ref{raichaid}) and (\ref{phi}) are solved numerically for the
following values of the parameters and the initial conditions: $m_\phi=
10^{-6} \, m_{\textsc{P}}$, $a_i=\sqrt{\gamma} \, \ell_{\textsc{P}}$, $\phi_i
= 3.6\times10^{11}j^{-15/2} m_{\textsc{P}}$ and  $\dot{\phi}_i = 2\times
10^{-6} \, m_{\textsc{P}}^2$, with $\dot{a}_i$ being fixed by Eq.~(\ref{H}).
This equation is also used to check the accuracy of the numerical solution.
The initial value for scalar field was chosen such that the uncertainty
principle \cite{tsujikawa-singh-maartens},
\begin{eqnarray}
|\phi_i\dot{\phi}_i| \gtrsim \frac{10^3}{j^{3/2}} \left( \frac{a_i}{a_*}
\right)^{12} m_{\textsc{P}}^3 \label{uncertainty-principle},
\end{eqnarray}
is marginally satisfied.

The ambiguity parameter $\ell$ is quantized, taking values
\begin{eqnarray}
\ell=1-\frac{1}{2n}; \quad n \in \mathbb{N}.
\end{eqnarray}
The other ambiguity parameter, $j$, takes half-integer values greater than
one. However, if one demands the initial value of the scalar field, $\phi_i$,
to be much smaller than the Planck mass (say, $\phi_i \lesssim 10^{-3}\,
m_{\textsc{P}}$), then the parameter $j$ is constrained to be much bigger than
one, namely, $j\gtrsim 87.3$. In what follows, we will consider $j$ to be a
continuously varying parameter with values greater or of the order of $100$.

As our numerical calculations show, after a short period of time, the
functions $D(q)$ e $F(q)$ approach their classical values, namely, $D=1$ and
$F=0$, and the semiclassical corrections can be neglected in
Eqs.~(\ref{H})--(\ref{phi}). The resulting simplified evolutionary equations
are then solved up to the end of the inflationary period.

\begin{figure}[t]
\includegraphics[width=8.1cm]{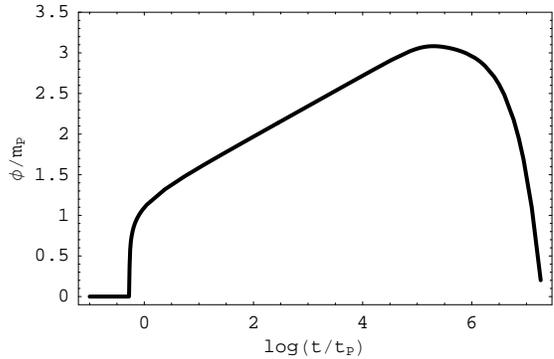}
\caption{Time evolution of the scalar field $\phi$ for the case $j=100$ and
$\ell=3/4$. Because of loop quantum effects, the scalar field increases
naturally from its initial value of about $3.6\times10^{-4}\, m_{\textsc{P}}$
to a maximum value of about $3\, m_{\textsc{P}}$, which guarantees a long
enough inflationary period. \label{fig:phi}}
\end{figure}

At the end of the semiclassical stage of evolution, the scalar field increases
from a small value (much smaller than the Planck mass) to a value of the order
of one Planck mass (see Fig.~\ref{fig:phi}). This is due to the fact that  the
first term on the right-hand-side of Eq.~(\ref{phi}) acts as an anti-friction
term, pushing the scalar field up the potential. During the subsequent
classical stage of evolution, the scalar field continues to increase for a
while, reaching a maximum value of about $3\, m_{\textsc{P}}$ (for $j=100$ and
$\ell =3/4$). This value of $\phi_{\mbox{\scriptsize{max}}}$ is enough for the
universe to expand about $60$ $e$-folds during the inflationary period.
Therefore, loop quantum effects can set the initial conditions for successful
chaotic inflation in a natural way \cite{tsujikawa-singh-maartens}.

At the end of the inflationary period, the scalar field begins to oscillate
around the minimum of the potential, transferring its energy to a radiation
fluid, thus reheating the universe. The decay of the scalar field into
radiation is governed by a dissipative coupling introduced in the evolutionary
equations, which now read
\begin{eqnarray}
\left( \frac{\dot{a}}{a} \right)^2 = \frac{8\pi}{3 m_{\textsc{P}}^2} \left[
\frac{\dot{\phi}^2}{2} + V(\phi)  + \rho_r \right], \label{H-reheating}
\end{eqnarray}
\vspace{-5mm}
\begin{eqnarray}
\frac{\ddot{a}}{a} = \frac{8\pi}{3 m_{\textsc{P}}^2} \left[ V(\phi) -
\dot{\phi}^2 -\rho_r \right], \label{raichaid-reheating}
\end{eqnarray}
\vspace{-5mm}
\begin{eqnarray}
\ddot{\phi} &=& -3\frac{\dot{a}}{a} \dot{\phi} - \frac{\partial V}{\partial
\phi}-\Gamma_\phi \dot{\phi}, \label{phi-reheating}
\end{eqnarray}
\vspace{-5mm}
\begin{eqnarray}
\dot{\rho_r} = -4\frac{\dot{a}}{a} \rho_r + \Gamma_\phi \dot{\phi}^2,
\label{rho-reheating}
\end{eqnarray}
where $\rho_r$ is the energy density of radiation and $\Gamma_\phi$ is the
dissipative coefficient. Since any preexisting radiation fluid would have been
diluted during the inflationary period, we choose the energy density of
radiation at the beginning of reheating to be zero, $\rho_{r,i}=0$. For the
dissipative coefficient we choose $\Gamma_\phi=2\times10^{-7}m_{\textsc{P}}$.

After a while, the energy density of the scalar field becomes much smaller
than the energy density of radiation, meaning that the former can be
consistently neglected. The evolutionary equations then become\footnote{In the
previous stages of evolution we have used the natural system of units, with
$\hbar=c=1$ and $m_{\textsc{P}}=G^{-1/2}=1.22\times10^{19}$ GeV, while here we
are using the international system of units.}
\begin{eqnarray}
\left( \frac{\dot{a}}{a} \right)^2 = \frac{8\pi G}{3c^2} \left[ \rho_{r,0}
\left( \frac{a_0}{a} \right)^4
+ \rho_{m,0} \left( \frac{a_0}{a} \right)^3 \right. \nonumber \\
+ \left. \rho_{de,0} \left( \frac{a_0}{a} \right)^{3(w+1)} \right],
\label{H-last}
\end{eqnarray}
\vspace{-5mm}
\begin{eqnarray}
\frac{\ddot{a}}{a} = -\frac{4\pi G}{3c^2}  \left[ 2 \rho_{r,0} \left(
\frac{a_0}{a} \right)^4 +
\rho_{m,0} \left( \frac{a_0}{a} \right)^3 \right. \nonumber \\
+ \left. (3w+1) \rho_{de,0} \left( \frac{a_0}{a} \right)^{3(w+1)} \right],
\label{raichaid-last}
\end{eqnarray}
where $a_0$ is today's value of the scale factor and
$\rho_{r,0}=4.13\times10^{-14}\mbox{ J/m}^3$,
$\rho_{m,0}=2.34\times10^{-10}\mbox{ J/m}^3$ and
$\rho_{de,0}=6.20\times10^{-10}\mbox{ J/m}^3$ are, respectively, today's
values of the energy density of radiation, matter, and dark energy. For these
values of $\rho_{r,0}$, $\rho_{m,0}$, and $\rho_{de,0}$, the value of the
Hubble constant is $H_0=71 \mbox{ km s}^{-1} \mbox{Mpc}^{-1}$. We take the
value $w=-1$ for the equation-of-state parameter of dark energy.

\begin{figure}[t]
\includegraphics[width=8.1cm]{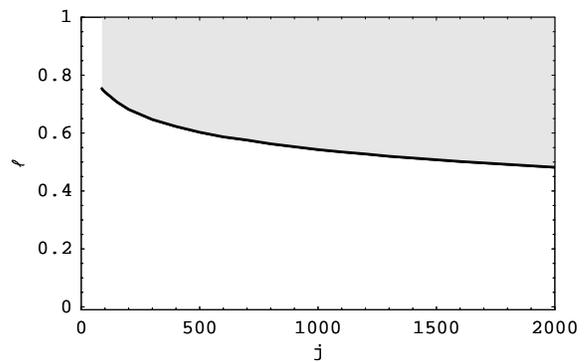}
\caption{The shaded region corresponds to values of the ambiguity parameters
$\ell$ and $j$ for which the universe expands at least $60$ $e$-folds during
the inflationary period. Such an expansion is achieved if the scalar field
grows from $\phi_i\ll\, m_{\textsc{P}}$ to $\phi_{\mbox{\scriptsize
max}}\gtrsim 3\, m_{\textsc{P}}$ during the pre-inflationary epoch. The
requirement that $\phi_i \lesssim 10^{-3} m_{\textsc{P}}$, together with the
choice $m_\phi=10^{-6}\, m_{\textsc{P}}$ and $\dot{\phi}_i = 2\times10^{-6}\,
m_{\textsc{P}}^2$, imposes the constraint $j \gtrsim 87.3$.
\label{fig:l-versus-j}}
\end{figure}

To finish this section, let us point out that for a given value of $j$, the
requirement that the universe expands enough during the inflationary period
(at least $60$ $e$-folds), imposes a lower bound on the value of $\ell$ (see
Fig.~\ref{fig:l-versus-j}).

\section{The gravitational-wave spectrum\label{sect-gravitational-waves}}

Gravitational waves are generated in an expanding universe, giving rise to a
spectrum extending over a wide range of frequencies \cite{grishchuk74,
starobinskii79, abbott-harari, Allen88, sahni, grishchuk-solokhin, allen97}.
In this section we calculate the full spectrum of the gravitational waves
generated within the loop quantum cosmology model described above.

Loop quantum effects introduce modifications not only to the dynamical
equations of evolution (\ref{H})--(\ref{phi}), but also to the equation for
tensor modes. However, in this paper, the latter will be neglected, allowing
for a considerable simplification of the calculations required to determine
the energy spectrum of gravitational waves, while keeping unchanged the main
features of the spectrum.

The tensor perturbations $h_{ij}$ to the Friedmann-Robertson-Walker metric,
\begin{eqnarray}
ds^{2}=a^2(\eta ) \left\{ -d\eta^{2} + \left[ \delta_{ij} + h_{ij} (\eta
,\textbf{x}) \right] dx^i dx^j \right\},
\end{eqnarray}
can be expanded in terms of plane waves
\begin{eqnarray}
h_{ij}(\eta,\textbf{x}) = \sqrt{8\pi G} \sum\limits_{p=1}^{2} \int
\frac{d^3k}{(2\pi)^{3/2} a(\eta)\sqrt{2k}} \nonumber \\
\times \left[ a_p(\eta,{\bf k}) \varepsilon_{ij}(\textbf{k},p)
e^{i\textbf{k}\cdot\textbf{x}} \xi(\eta,k) + \mbox{H.c.}\right],
\end{eqnarray}
where $G$ is the gravitational constant, $p$ runs over the two polarizations
of the gravitational waves, $k=|\textbf{k}|=2\pi a/\lambda =a\omega $ is the
co-moving wave number, $a_p$ is the annihilation operator, and
$\varepsilon_{ij}$ is the polarization tensor. The mode function $\xi(\eta,k)$
obeys the equation of a parametric oscillator,
\begin{eqnarray}
\xi^{\prime\prime}+ \left( k^2-U \right) \xi = 0, \label{tensor_perturb}
\end{eqnarray}
where the potential $U$ is given by
\begin{eqnarray}
U = \frac{a^{\prime\prime}}{a}
\end{eqnarray}
and a prime denotes a derivative with respect to conformal time $\eta$.

For $k^2 \lesssim U$, the above equation describes the production of
gravitational waves with frequency $\omega=k/a$, while for $k^2 \gg U$ the
equation is that of an harmonic oscillator, implying that no gravitational
waves are produced.

\begin{figure}[t]
\includegraphics[width=8.1cm]{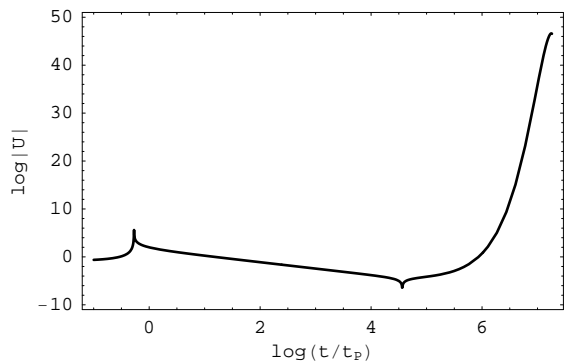}
\caption{Potential $U$ as a function of (cosmic) time $t$ for the case
$\ell=3/4$ and $j=100$. The inflationary barrier is located on the right (for
$t\gtrsim 10^6 t_{\textsc{P}}$), while the much smaller barrier arising due to
loop quantum effects is located on the left (for $t\lesssim t_{\textsc{P}}$).
\label{fig-U}}
\end{figure}

Within the standard model of cosmology, the potential $U$ has a pronounced
barrier at the time of inflation, giving rise to a copious production of
gravitational waves with frequencies up to the gigahertz. Within loop quantum
cosmology, besides the above mentioned inflationary barrier, the potential $U$
has another barrier at very early times (see Fig.~\ref{fig-U}). The presence
of this barrier leads to the creation of extra gravitational waves of low
frequency. This barrier is slightly more pronounced, if one takes into account
loop quantum corrections to the equation of tensor perturbations.

In order to calculate the energy spectrum of the cosmological gravitational
waves generated during the evolution of the universe, from the semiclassical
pre-inflationary stage of evolution till the present time, we use the method
of continuous Bogoliubov coefficients. This method, first applied by Parker
\cite{parker} to particle production in an expanding universe and then
extended to the case of gravitons
\cite{henriques94,henriques-moorhouse-mendes,mendes-henriques-moorhouse}, can
be summarized as follows (for applications of this method to several
cosmological models see
Refs.~\cite{henriques04,sa-henriques1,sa-henriques2,sa-henriques-potting,sa-henriques3}).

The gravitational-wave spectral energy density parameter,
$\Omega_\textsc{gw}$, is defined as
\begin{eqnarray}
\Omega_\textsc{gw} = \frac{8\hbar G}{3\pi c^5 H_0^2} \omega_0^4 \beta_0^2,
\label{sedp}
\end{eqnarray}
where $\hbar$ is the reduced Planck constant, $G$ is the gravitational
constant, $c$ is the speed of light, $H$ is the Hubble parameter, $\omega$ is
the gravitational-wave frequency, $\beta$ is a Bogoliubov coefficient and the
subscript $0$ denotes quantities evaluated at the present time.

The angular frequency $\omega_0$ takes values ranging from about
$1.4\times10^{-17}\mbox{ rad/s}$ (corresponding to a wavelength equal, today,
to the Hubble distance) to about $10^{9}\mbox{ rad/s}$ (corresponding to a
wavelength equal to the Hubble distance at the end of the inflationary
period).

The number of gravitons at a certain time is given by the squared Bogoliubov
coefficient,
\begin{eqnarray}
|\beta(t)|^2 = \frac14 [X(t)-Y(t)] [X^*(t)-Y^*(t)],
\end{eqnarray}
where $*$ denotes complex conjugate and the functions $X$ and $Y$ are
solutions of the system of differential equations
\begin{eqnarray}
\dot{X} = - i \frac{\omega_0 a_0}{a} Y, \label{XX}
\end{eqnarray}
\vspace{-5mm}
\begin{eqnarray}
\dot{Y} = -i \frac{a}{\omega_0 a_0} \left[ \left( \frac{\omega_0 a_0}{a}
\right)^2- \frac{\ddot{a}}{a}- \left( \frac{\dot{a}}{a} \right)^2\right]X,
\label{YY}
\end{eqnarray}
which is integrated with initial conditions $X(t_{i})=Y(t_{i})=1$,
corresponding to the absence of gravitons at the beginning of the evolution.
In the above system of differential equations, the scale factor, $a(t)$, and
its first and second derivatives, $\dot{a}(t)$ and $\ddot{a}(t)$, are
determined from the evolutionary equations presented in the previous section,
namely, Eqs.~(\ref{H})--(\ref{phi}) for the semiclassical stage of evolution
and the inflationary period , (\ref{H-reheating})--(\ref{rho-reheating}) for
the reheating period, and (\ref{H-last})--(\ref{raichaid-last}) for the
radiation-dominated, matter-dominated and dark energy-dominated periods.

Using the above outlined method of continuous Bogoliubov coefficients, we can
calculate the gravitational-wave spectrum for different values of the
ambiguity parameters $j$ and $\ell$. Let us first analyze the case of fixed
$j$ (say, $j=100$) and varying $\ell$.

\begin{figure}[t]
\includegraphics[width=8.1cm]{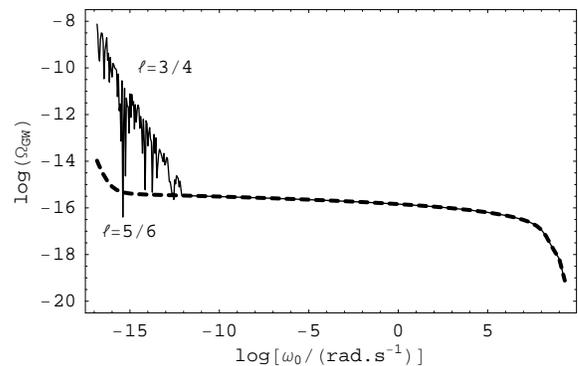}
\caption{Gravitational-wave spectra for $\ell=3/4$ and $\ell=5/6$ (dashed
line). In both cases $j=100$.  \label{fig:espectroJ100L34}}
\end{figure}

\begin{figure}[t]
\includegraphics[width=8.1cm]{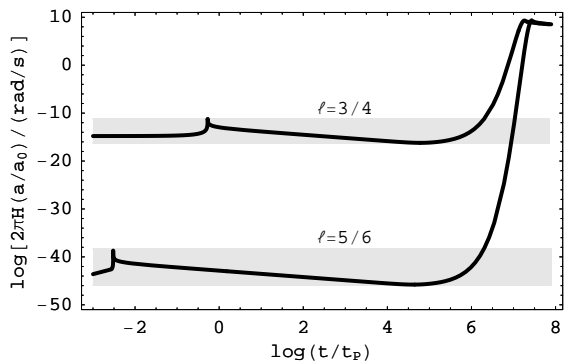}
\caption{Today's frequency of a gravitational wave as a function of the time
at which this wave first crossed the Hubble horizon. For $\ell=3/4$,
gravitational waves generated in the early universe have frequencies, today,
of the order of $(10^{-11}-10^{-17})\mbox{ rad/s}$ (shaded region),
corresponding to wavelengths of the order or smaller than the Hubble horizon
today. For $\ell=5/6$, gravitational waves generated in the early universe
have, today, frequencies of the order of $(10^{-38}-10^{-45})\mbox{ rad/s}$,
corresponding to wavelengths much bigger than the Hubble horizon today. For
both cases today's maximum frequency of the gravitational waves is of the
order of $10^9\mbox{ rad/s}$, corresponding to waves that crossed the Hubble
horizon at the end of the inflationary period ($t\sim 3\times10^7\,
t_\textsc{P}$). \label{fig:freq_todos}}
\end{figure}

For $\ell=3/4$, loop quantum effects leave a clear signature on the spectrum,
namely, an over-production of low-frequency gravitational waves (see
Fig.~\ref{fig:espectroJ100L34}). For $\ell=5/6$ the spectrum shows no
influence of these effects. This can be understood as follows. Today's
frequency of a gravitational wave that crossed the Hubble horizon at time $t$
is given by $2\pi H(t) [a(t)/a(t_0)]$. For $\ell=3/4$, gravitational waves
generated in the early universe ($t\lesssim10^5\, t_\textsc{P}$) have
frequencies, today, of the order of $(10^{-11}-10^{-17})\mbox{ rad/s}$,
corresponding to wavelengths of the order or smaller than the Hubble horizon
today (see Fig.~\ref{fig:freq_todos}). Therefore, these gravitational waves
leave their imprint on the spectrum at low frequencies. For $\ell=5/6$,
gravitational waves generated in the early universe have, today, frequencies
of the order of $(10^{-38}-10^{-45})\mbox{ rad/s}$, corresponding to
wavelengths much bigger than the Hubble horizon today. Therefore, these
gravitational waves leave no imprint on the spectrum\footnote{The small rise
on the spectrum at low frequencies for $\ell=5/6$ is due to another effect,
namely, an extra production of gravitational waves in recent epochs, when the
universe became matter dominated and, then, dark-energy dominated.}. Note that
in the case $\ell=3/4$ the scale factor grows about $60$ $e$-folds during the
inflationary period, while in the case $\ell=5/6$ the growth of the scale
factor is about $140$ $e$-folds.

\begin{figure}[t]
\includegraphics[width=8.1cm]{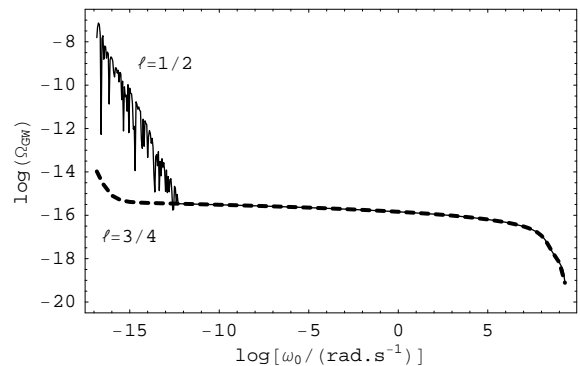}
\caption{Gravitational-wave spectra for $\ell=1/2$ and $\ell=3/4$ (dashed
line). In both cases $j=2000$.  \label{fig:espectroJ2000L12}}
\end{figure}

For the case $j=2000$ the situation is similar to the one described above (see
Fig.~\ref{fig:espectroJ2000L12}). Loop quantum effects leave an imprint on the
gravitational-wave spectrum only if the ambiguity parameter $\ell$ takes such
a value ($\ell=1/2$) that the universe expands about $60$ $e$-folds during the
inflationary period. If the universe expands much more than $60$ $e$-folds
(for instance, in the case $\ell=3/4$ the scale grows about $240$ $e$-folds),
the gravitational waves generated before the inflationary period have
wavelengths, today, much bigger than the Hubble horizon and, consequently,
they leave no imprint on the gravitational-wave energy spectrum.

We have been assuming that the ambiguity parameter $\ell$ is quantized, taking
values $\ell=1-1/(2n)$, with $n\in \mathbb{N}$. If, however, we consider
$\ell$ to be a free parameter, changing continuously from $0$ to $1$, then our
conclusions need to be slightly adapted. Namely, for each value of $j$ there
is a critical value of the ambiguity parameter, $\ell_{\mbox{\scriptsize
crit}}$, for which the universe expands $60$ $e$-folds during the inflationary
period, the minimum required in standard inflationary cosmology. For values of
$\ell$ close to $\ell_{\mbox{\scriptsize crit}}$, the gravitational waves
generated during the pre-inflationary epoch leave an imprint on the spectrum
at low frequencies. As we (continuously) increase $\ell$, this imprint shows
up at lower and lower frequencies, completely disappearing when today's
frequency of the generated waves is so low ($\omega_0 \lesssim 10^{-17}\mbox{
rad/s}$), that it corresponds to a wavelength greater than the Hubble horizon.
On the contrary, if we consider values of $\ell$ smaller than the critical one
(in which case the scale factor does not grow enough during the inflationary
period), the imprint of the pre-inflationary gravitational waves appears on
the spectrum at higher frequencies.

\begin{figure}[t]
\includegraphics[width=8.1cm]{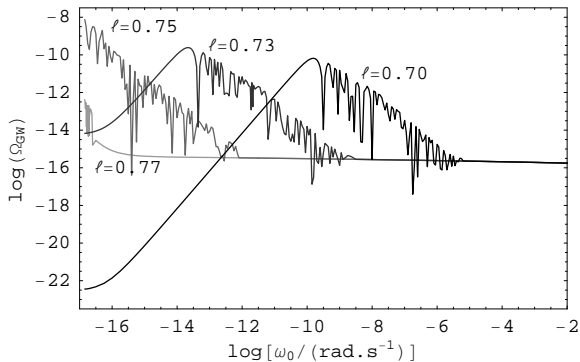}
\caption{Gravitational-wave spectra for $\ell=0.70$, $0.73$, $0.75$ and $0.77$
($j=100$ in all cases). Loop quantum effects leave a signature on the spectra,
namely, an over-production of gravitational waves at low frequencies. As
$\ell$ increases this signature is located at increasingly lower frequencies.
For $\ell=0.77$ the signature moved to frequencies so low, that it becomes
almost unnoticeable in the spectrum. \label{fig:espectroJ100L073}}
\end{figure}

The above conclusions are illustrated in Fig.~\ref{fig:espectroJ100L073}. For
$j=100$ the critical value of the ambiguity parameter $\ell$ is about $0.75$
(see Fig.~\ref{fig:l-versus-j}). In this case, loop quantum effects leave
their signature on the spectrum at frequencies $\omega_0\lesssim
10^{-12}\mbox{ rad/s}$. For $\ell=0.70$ and $\ell=0.73$ the gravitational
waves generated prior to the inflationary period leave an imprint at
frequencies higher than in the critical case, $\omega_0\lesssim 10^{-5}\mbox{
rad/s}$ and $\omega_0\lesssim 10^{-8}\mbox{ rad/s}$, respectively. However,
for such values of the ambiguity parameters, the scale factor grows about $45$
and $54$ $e$-folds, respectively, during the inflationary period, which is
less than required by standard inflationary cosmology. For the $\ell=0.77$
(above $\ell_{\mbox{\scriptsize crit}}=0.75$) the signature of loop quantum
cosmology is located at frequencies lower than in the critical case,
$\omega_0\lesssim 10^{-16}\mbox{ rad/s}$. In this case, the scale factor grows
about $71$ $e$-folds during the inflationary period.

\begin{figure}[t]
\includegraphics[width=8.1cm]{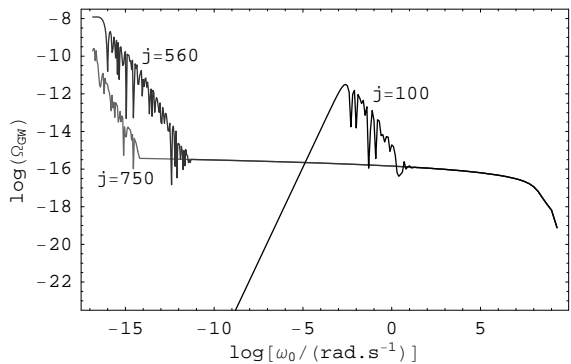}
\caption{Gravitational-wave spectra for $j=100$, $560$ and $750$ ($\ell=0.6$
in all cases). As $j$ increases the loop quantum signature shows up at
increasingly lower frequencies. \label{fig:espectroL060}}
\end{figure}

To conclude, let us compare gravitational-wave spectra for a fixed value of
$\ell$ (say, $\ell=0.6$) and varying $j$. For $\ell=0.6$, enough growth of the
scale factor during the inflationary period is guaranteed for $j \gtrsim 560$
(see Fig.~\ref{fig:l-versus-j}). Therefore, as expected, for $j=560$ loop
quantum effects leave their imprint at low frequencies, $\omega_0\lesssim
10^{-11}\mbox{ rad/s}$ (see Fig.~\ref{fig:espectroL060}). For $j=750$ the loop
quantum signature moves to lower frequencies, $\omega_0\lesssim 10^{-14}\mbox{
rad/s}$, while for $j=100$ it moves to higher frequencies, $\omega_0\lesssim
10\mbox{ rad/s}$. In the latter case, the imprint of loop quantum cosmology
lies in the frequency band accessible to the Laser Interferometer Space
Antenna (LISA) and, in principle, could be seen by this detector. Note,
however, that the growth of the scale factor during the inflationary period in
this case is just about $28$ $e$-folds, which is manifestly insufficient
within standard inflationary cosmology.

To finish this section, let us point out that measurements of the cosmic
microwave background radiation can be used to derive an upper limit on the
gravitational-wave spectral energy density parameter, namely,
$\Omega_{\textsc{gw}} < 1.4 \times 10^{-10}$ for
$\omega_0=1.4\times10^{-17}\mbox{ rad/s}$ \cite{allen97}. Some
gravitational-wave spectra shown in this paper do not satisfy this bound.
Taking into account that the inflaton mass $m_\phi$ determines the overall
vertical displacement of the gravitational-wave spectrum, one just needs to
consider lower values of $m_\phi$ in order to make these spectra compatible
with the above mentioned upper limit.

\section{Conclusions\label{sect-conclusions}}

In this work we have investigated the generation of gravitational waves within
loop quantum cosmology models. For such models, the evolution of the universe
is naturally divided in two stages, corresponding to the semiclassical and the
classical regimes. In the former, loop quantum effects introduce modifications
to the dynamical equations describing the evolution of the early universe,
while in the latter the evolution proceeds according to the usual general
relativity equations. The transition between the two regimes takes place at
very early times.

For the semiclassical regime we have assumed that the corrections to the
dynamical equations are of the inverse-volume type, leaving for a future
investigation the holonomy corrections. Inverse-volume corrections involve two
ambiguity parameters, $j$ and $\ell$, which we have assumed to be free
parameters.

For the classical regime, we have assumed that the evolution of the universe
proceeds according to the usual standard inflationary model, i.e., a
inflationary epoch (of the chaotic type) is followed by reheating and then,
successively, by radiation-dominated, matter-dominated and dark
energy-dominated periods of evolution.

We have also assumed that, initially, the inflaton field $\phi$ is located
near the origin of the chaotic-type potential, taking values much smaller than
the Planck mass. This choice of initial conditions is admissible, since loop
quantum corrections to the dynamical equations guarantee that the scalar field
$\phi$ is pushed up the potential, reaching maximum values of the order of the
Planck mass at the beginning of the inflationary period (see
Fig.~\ref{fig:phi}), which is required to obtain enough inflation.

The assumption that $\phi_i \ll m_{\textsc{P}}$ and $\dot{\phi}_i\ll
m_{\textsc{P}}^2$, together with the requirement that initially the scalar
field satisfies marginally the uncertainty principle, imposes a lower bound on
the value of the ambiguity parameter $j$, namely, $j\gtrsim100$. Another
constraint on the values of $j$ and $\ell$ comes from the requirement that the
scale factor grows at least $60$ $e$-folds during the inflationary period. Our
numerical calculations show that this condition is satisfied just for the
values of the ambiguity parameters $j$ and $\ell$ corresponding to the shaded
region of Fig.~\ref{fig:l-versus-j}.

Loop quantum effects introduce modifications not only to the dynamical
equations describing the evolution of the universe, but also to the equation
of tensor modes. However, in this paper, we have taken into account only the
modifications to the dynamical equations, thus simplifying considerably the
calculations without loosing the main features of the gravitational-wave
energy spectrum.

To calculate the full gravitational-wave energy spectrum we have used the
method of continuous Bogoliubov coefficients. Our analysis shows that, for
certain conditions, loop quantum effects leave a clear signature on the
spectrum, namely, an over-production of low-frequency gravitational waves.
This signature is present on the gravitational-wave spectrum only if the
growth of the scale factor during the inflationary period does not exceed
significantly the minimum growth required in standard inflationary cosmology,
namely, $60$ $e$-folds. If the scale factor grows much more than this value,
gravitational waves generated prior to the inflationary period have, today, a
wavelength much bigger than the Hubble horizon, leaving no imprint on the
gravitational-wave spectrum. On the other hand, if the growth of the scale
factor during the inflationary period is smaller than $60$ $e$-folds, then the
imprint of loop quantum cosmology is clearly seen on the gravitational-wave
spectrum, at frequencies which increase with decreasing number of $e$-folds of
expansion during inflation. For example, in the case $\ell=0.6$ and $j=100$,
for which about $28$ $e$-folds of expansion are obtained during inflation,
loop quantum effects leave their signature at LISA frequency band. Taken into
account the above comments, we conclude that the values of the ambiguity
parameters $j$ and $\ell$ for which a signature of loop quantum cosmology
shows up on the gravitational-wave spectrum are those corresponding to a
narrow band around the thick line of Fig.~\ref{fig:l-versus-j}, i.e., these
values for which the scale factor grows about $60$ $e$-folds during the
inflationary period.

Our results shows that, contrary to what is usually assumed, inflation does
not necessarily erase all the information on physical features present in the
pre-inflationary era. Indeed, as we have shown, within loop quantum cosmology
physical processes taking place in the very early universe, prior to the
inflationary period, leave their imprint on the spectrum of the gravitational
waves at very low frequencies, corresponding to wavelengths of the order of
the Hubble distance. Despite the fact that the gravitational-wave spectral
energy density parameter $\Omega_\textsc{gw}$ for such frequencies may be
quite high, a direct detection is not possible. Nevertheless, these
gravitational waves may have left their imprint on the cosmic microwave
background radiation and on large scale structures, in which case they will
allow us to test present-day theories about the quantum origin of the
universe.

\begin{acknowledgments}
This work was supported in part by the Funda\c{c}\~ao para a Ci\^encia e a
Tecnologia, Portugal.
\end{acknowledgments}


\newpage

\begin{thebibliography}{99}

\bibitem{rovelli-2008} C.~Rovelli, Living Rev.\ Relativity \textbf{11}, 5 (2008).

\bibitem{bojowald-varios} M.~Bojowald, Class.\ Quantum Grav.\ \textbf{17}, 1489 (2000);
\textbf{17}, 1509 (2000); \textbf{18}, 1055 (2001); \textbf{18}, 1071 (2001);
\textbf{18}, L109 (2001); \textbf{19}, 2717 (2002); \textbf{19}, 5113 (2002);
\textbf{20}, 2595 (2003); Phys.\ Rev.\ Lett.\ \textbf{86}, 5227 (2001);
\textbf{87}, 121301 (2001); \textbf{89}, 261301 (2002); Phys.\ Rev.\ D
\textbf{64}, 084018 (2001); Adv.\ Theor.\ Math.\ Phys.\ \textbf{7}, 233 (2003).

\bibitem{bojowald-2008} M.~Bojowald, Living Rev.\ Relativity \textbf{11}, 4
(2008).

\bibitem{bojowald-lidsey-mulryne-singh-tavakol} M.~Bojowald, J.~E.~Lidsey,
D.~J.~Mulryne, P.~Singh, and R.~Tavakol, Phys.\ Rev.\ D \textbf{70}, 043530
(2004).

\bibitem{tsujikawa-singh-maartens} S.~Tsujikawa, P.~Singh, and R.~Maartens,
Class.\ Quantum Grav.\ \textbf{21}, 5767 (2004).

\bibitem{hossain} G.~M.~Hossain, Class.\ Quantum Grav.\ \textbf{22}, 2511
(2005).

\bibitem{copeland-mulryne-nunes-shaeri} E.~J.~Copeland, D.~J.~Mulryne,
N.~J.~Nunes, and M.~Shaeri, Phys.\ Rev.\ D \textbf{77}, 023510 (2008).

\bibitem{bojowald-calcagni} M.~Bojowald and G.~Calcagni, JCAP \textbf{03} 032
(2011).

\bibitem{bojowald-calcagni-tsujikawa-A} M.~Bojowald, G.~Calcagni, and
S.~Tsujikawa, ``Observa\-tional constraints on loop quantum cosmology",
arXiv:1101.5391 [astro-ph.CO].

\bibitem{bojowald-calcagni-tsujikawa-B} M.~Bojowald, G.~Calcagni, and
S.~Tsujikawa, ``Observa\-tional test of inflation in loop quantum cosmology",
arXiv:1107.1540 [gr-qc].

\bibitem{mielczarek-szydlowski} J.~Mielczarek and
M.~Szydlowski, Phys.\ Lett.\ B \textbf{657}, 20 (2007).

\bibitem{grain-et-al} J.~Grain, A.~Barrau, and A.~Gorecki,
Phys.\ Rev.\ D \textbf{79}, 084015 (2009).

\bibitem{bojowald-2004} M.~Bojowald, Pramana \textbf{63}, 765 (2004).

\bibitem{meissner} K.~A.~Meissner, Class.\ Quantum Grav.\ \textbf{21}, 5245 (2004).

\bibitem{grishchuk74} L.~P.~Grishchuk, Sov.\ Phys.\ JETP \textbf{40}, 409
(1974).

\bibitem{starobinskii79} A.~A.~Starobinskii, JETP Lett.\ \textbf{30}, 682 (1979).

\bibitem{abbott-harari} L.~F.~Abbott and D.~D.~Harari, Nucl. Phys. B \textbf{264}, 487 (1986).

\bibitem{Allen88} B.~Allen, Phys.\ Rev.\ D \textbf{37}, 2078 (1988).

\bibitem{sahni} V.~Sahni, Phys.\ Rev.\ D \textbf{42}, 453 (1990).

\bibitem{grishchuk-solokhin} L.~P.~Grishchuk and M.~Solokhin,
Phys.\ Rev.\ D \textbf{43}, 2566 (1991).

\bibitem{allen97} B.~Allen, in \emph{Proceedings of the Les Houches School
on Astrophysical Sources of Gravitational Waves (Les Houches, France, 1995)},
edited by J.-A. Marck and J.-P. Lasota (Cambridge University Press, Cambridge,
England, 1997), p.~373.

\bibitem{parker} L.~Parker, Phys.\ Rev.\ \textbf{183}, 1057 (1969).

\bibitem{henriques94} A.~B.~Henriques, Phys.\ Rev.\ D \textbf{49},
1771 (1994).

\bibitem{henriques-moorhouse-mendes}  R.~G.~Moorhouse, A.~B.~Henriques,
and L.~E.~Mendes, Phys.\ Rev.\ D \textbf{50}, 2600 (1994).

\bibitem{mendes-henriques-moorhouse} L.~E.~Mendes, A.~B.~Henriques,
and R.~G.~Moorhouse, Phys.\ Rev.\ D \textbf{52}, 2083 (1995).

\bibitem{henriques04} A.~B.~Henriques, Class.\ Quantum Grav. \textbf{21},
3057 (2004); \textbf{24}, 6431(E) (2007).

\bibitem{sa-henriques1} P.~M.~S\'a and A.~B.~Henriques,
Phys.\ Rev.\ D \textbf{77}, 064002 (2008).

\bibitem{sa-henriques2} P.~M.~S\'a and A.~B.~Henriques,
Gen.\ Relativ.\ Gravit. \textbf{41}, 2345 (2009).

\bibitem{sa-henriques-potting} A.~B.~Henriques, R.~Potting, and P.~M.~S\'a,
Phys.\ Rev.\ D \textbf{79}, 103522 (2009).

\bibitem{sa-henriques3} P.~M.~S\'a and A.~B.~Henriques,
Phys.\ Rev.\ D \textbf{81}, 124043 (2010).

\end{thebibliography}
\end{document}